\documentclass[9pt]{extarticle}
\usepackage{spconf,amsmath,graphicx}
\usepackage{amssymb}
\usepackage{multirow}
\usepackage{booktabs}
\usepackage{ctable}

\title{FRCRN: Boosting Feature Representation using Frequency Recurrence for Monaural Speech Enhancement}
\twoauthors
  {Shengkui Zhao, Bin Ma}
	{Alibaba Group\\
	\{shengkui.zhao, b.ma\}@alibaba-inc.com\\}
  {Karn N. Watcharasupat, Woon-Seng Gan\sthanks{This work was jointly supported by the Alibaba-NTU Singapore Joint Research Institute via the Alibaba Innovative Research (AIR) Program (Ref. AN-GC-2020-015).}}
	{School of Electrical and Electronic Engineering\\
	Nanyang Technological University (NTU), Singapore\\
	\{karn001, ewsgan\}@ntu.edu.sg}
\begin{document}
%
\maketitle
\begin{abstract}
Convolutional recurrent networks (CRN) integrating a convolutional encoder-decoder (CED) structure and a recurrent structure have achieved promising performance for monaural speech enhancement. However, feature representation across frequency context is highly constrained due to limited receptive fields in the convolutions of CED. In this paper, we propose a convolutional recurrent encoder-decoder (CRED) structure to boost feature representation along the frequency axis. The CRED applies frequency recurrence on 3D convolutional feature maps along the frequency axis following each convolution, therefore, it is capable of catching long-range frequency correlations and enhancing feature representations of speech inputs. The proposed frequency recurrence is realized efficiently using a feedforward sequential memory network (FSMN). Besides the CRED, we insert two stacked FSMN layers between the encoder and the decoder to model further temporal dynamics. We name the proposed framework as Frequency Recurrent CRN (FRCRN). We design FRCRN to predict complex Ideal Ratio Mask (cIRM) in complex-valued domain and optimize FRCRN using both time-frequency-domain and time-domain losses. Our proposed approach achieved state-of-the-art performance on wideband benchmark datasets and achieved 2nd place for the real-time fullband track in terms of Mean Opinion Score (MOS) and Word Accuracy (WAcc) in the ICASSP 2022 Deep Noise Suppression (DNS) challenge.
\end{abstract}
\section{Introduction}
\label{sec:intro}

The target speech is often severely polluted by the additive background noise and reverberations in speech communication and automatic speech recognition (ASR) applications. The corrupted speech causes a reduction in speech perceptual quality and intelligibility as well as the automatic speech recognition (ASR) performance. The goal of speech enhancement is to extract the signal of interest from the corrupted speech for better perceptual quality and intelligibility as well as a more robust speech recognition performance.

Monaural speech enhancement has been considered a challenging problem for decades. Recently, deep learning-based methods have made significant progress
with the simulated noisy speech as input and the clean speech as target. Many deep models of different structures have been studied for speech enhancement, such as feedforward neural networks (FNN) \cite{Gao2016}, recurrent neural networks (RNN) \cite{Gao2018} and convolutional neural networks (CNN) \cite{Park2017}. The FNN model \cite{Gao2016} performs on a short context window and cannot leverage long-term contexts of speech signals. The RNN model \cite{Gao2018} can handle long-term contexts in a sequence-based manner, but often require high-level handcrafted features such as MFCC. The CNN model \cite{Park2017} is able to extract high-level features, but mostly focuses on local temporal-spectral patterns.
By leveraging both CNN and RNN, convolutional recurrent networks (CRN) were introduced to speech enhancement \cite{Tan2018, Zhao2018}. The CRN integrates a convolutional encoder-decoder (CED) structure and a recurrent structure. 
In CED, the encoder extracts high-level features from local temporal-spectral patterns and the decoder reconstructs the target map. The recurrent structure fed with the high-level features from the encoder further models the long-term temporal dependencies. 
Therefore, it is able to extract high-level features by the CED structure and models long-term temporal dependencies by the recurrent structure. 
The CRN has been shown to be very effective for speech enhancement \cite{Tan2018, Zhao2018} and the extended complex-valued  DCCRN  \cite{Hu2020} achieved the best performance in the Interspeech 2020 DNS Challenge. In \cite{Zhao2021}, it was demonstrated that CRN heavily relies on the representation power of the convolutions in CED and a complex convolutional block attention module (CCBAM) was introduced to boost the feature representation, resulting in improved performance.
 However, due to the limited receptive fields of convolutions, CRN still cannot capture well the long-range correlations along the frequency axis.

In this work, we are motivated by the frequency correlation modelling study in \cite{Lv2021} and propose a novel convolutional recurrent encoder-decoder (CRED) structure to boost feature representation along the frequency axis. Different from the pure convolution operations in CED, we add a frequency recurrence after each convolution in CRED. The frequency recurrence is applied on the 3D convolutional feature maps along the frequency axis. Specifically, on each time frame, a frequency sequence is first formed from lower frequencies to upper frequencies with the channels axis as the ``feature" dimension. Then, the frequency sequences are transformed by a recurrent network implemented by the feedforward sequential memory networks (FSMN) \cite{Zhang2018}.  The convolutional layer and the frequency recurrent layer form a convolutional recurrent (CR) block. We form CRED by stacking multiple CR blocks in both encoder and decoder. CRED is expected to capture not only the local temporal-spectral structures but also the long-range frequency dependencies. By further adding a time-recurrent structure as CRN, we form a new framework called Frequency Recurrent CRN (FRCRN). Unlike the previous work \cite{Lv2021} focusing on modelling the temporal dependency in the time-recurrent structure, our work focuses on improving the overall feature representation of the encoder-decoder structure. 
\begin{figure*}[h]
  \centering
  \includegraphics[width=15cm]{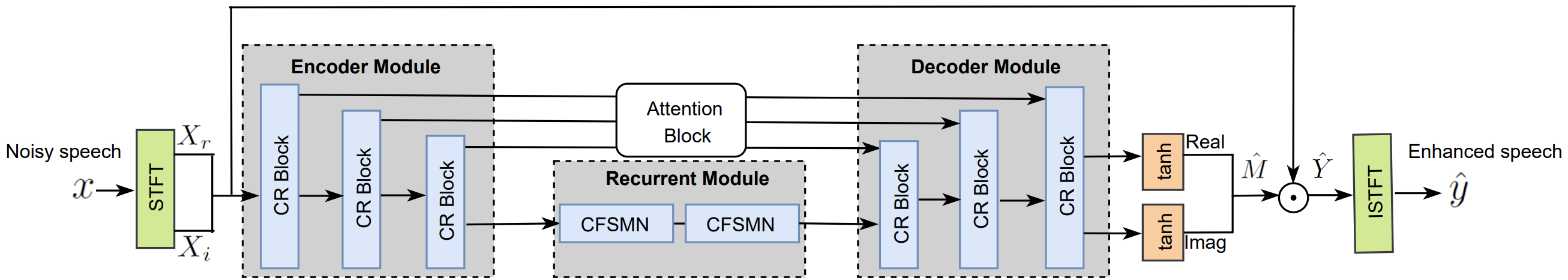}
  \caption{Network architecture of the proposed FRCRN model. (The symbol $\odot$ stands for element-wise complex multiplication.)}
\end{figure*}

To further improve performance, we implement the FRCRN in complex-valued operations and predict the complex Ideal Ratio Mask (cIRM) \cite{Williamson2015}. Moreover, we perform a joint optimization using both time-frequency domain and time-domain loss functions. Our experimental results show that the FRCRN model performs well for both wideband and fullband speech signals. The proposed FRCRN model achieves state-of-the-art (SOTA) results on the DNS-2020 dataset \cite{Reddy2020} and the Voicebank+Demand dataset \cite{Botinhao2016}. Our submission to the ICASSP 2022 Deep Noise Suppression (DNS) challenge (DNS-2022) \cite{Dubey2022} ranked overall 2nd place for the real-time fullband nono-personalized track in terms of Mean Opinion Score (MOS) and Word Accuracy (WAcc).

\section{The Proposed FRCRN Model}
\subsection{The overall architecture}
The overall architecture of the proposed FRCRN model is illustrated in Fig. 1. Our purpose is to estimate the clean speech signal $y$ from the corrupted speech signal $x=y \ast h+z \in \mathbb{R}$ where the time index is omitted for simplicity. The corruption process is made by a reverberation operation $y \ast h$ and an additive noise $z$. The signal $x$ is used as an input and first transformed to time-frequency spectrogram by applying the short-time Fourier transform (STFT). And then it is sent to the FRCRN model to predict the cIRM target. The \emph{tanh} activation function is applied to bound the estimates by $[-1,1]$. The enhanced spectrogram $\hat{Y}$ is obtained by multiplying the cIRM estimate $\hat{M}$ with the noisy spectrogram $X$.  The time-domain estimate $\hat{y}$ is obtained by applying ISTFT on $\hat{Y}$. Our FRCRN model is mainly comprised of the proposed CRED and a recurrent module. The CRED comprises an encoder module and a decoder module. Both modules comprise multiple convolutional recurrent (CR) blocks which will be described in the following section. To ensure the output has the same shape as the input, the CRED architecture is symmetric. The recurrent module comprises two stacked complex FSMN (CFSMN) layers. In FRCRN, the encoder extracts high-level feature representations and the decoder reconstructs the target map. The recurrent module models the long-term temporal dependencies.
The skip connections facilitate optimization by connecting each block in the encoder to its corresponding block in the decoder. We further add the attention block CCBAM \cite{Zhao2021} on the skip pathway to facilitate information flow. Note that we take all convolutions to be causal in time by applying asymmetrical paddings. 
\begin{figure*}[ht]
  \centering
  \includegraphics[width=16cm]{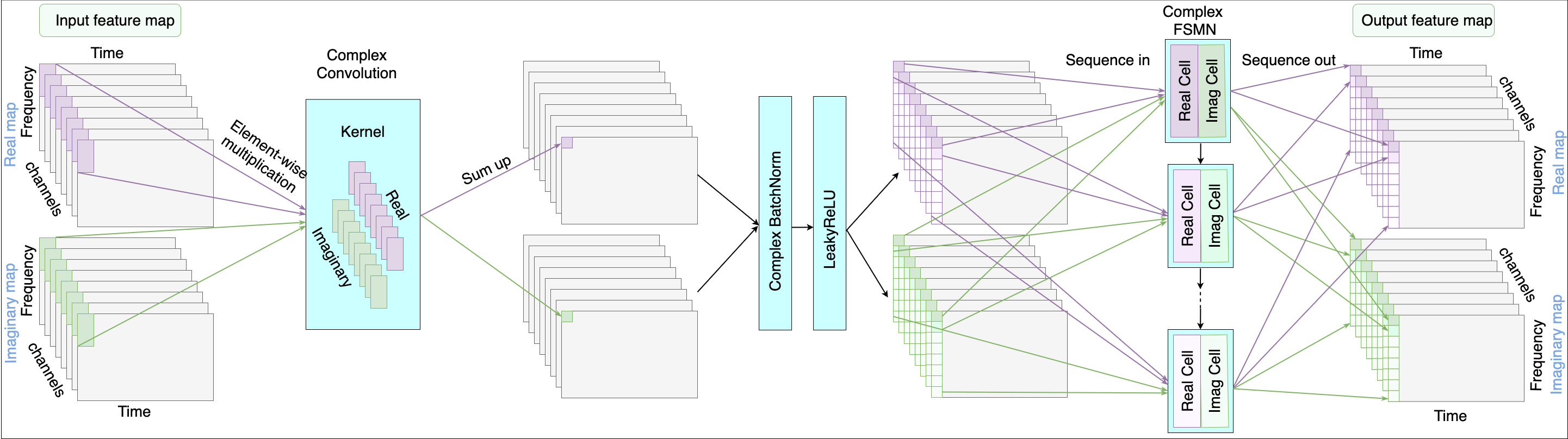}
  \caption{Network architecture of the CR block. Each CR block comprises a complex Conv2d, a complex BN, a LeakyReLU, and a CFSMN.}
  \label{fig2}
\end{figure*}

\subsection{Convolutional recurrent (CR) block}
The detailed architecture of the CR block is shown in Fig. 2. Each CR block comprises a complex 2D convolution layer (Conv2d), a complex batch normalization (BN), a LeakyReLU function, and a CFSMN layer. The operation of complex Conv2d is given as follows. Let the complex 3D input feature matrix be $V=V_r+jV_i\in \mathbb{C}^{C\times T\times F}$, where $V_r$ is the real part and $V_i$ the imaginary part. $C, T, F$ denote channel, frame and frequency dimensions, respectively.  Let the convolutional kernel be $W=W_r+jW_i\in \mathbb{C}^{C^\prime\times T^\prime\times F^\prime}$, where $C^\prime$ denotes the number of kernels and $T^\prime\times F^\prime$ denotes the kernel size. The output of the complex Conv2d $U=U_r+jU_i\in \mathbb{C}^{C^\prime\times T\times F^{\prime\prime}}$ can be formulated as:
\begin{align}
U_r = V_r \otimes W_r - V_i \otimes W_i \nonumber \\
U_i = V_r \otimes W_i + V_i \otimes W_r
\end{align}
where $\otimes$ stands for real-valued convolutional filtering. Here, we perform causal convolutions with $T^\prime=2$, $\mathrm{\emph{stride}}=1$ and using zero-padding on the time direction. On the frequency direction, we use $F^\prime=5$, $\mathrm{\emph{stride}}=2$ without zero-padding, which halves feature maps in frequency axis in the encoder block by block. For all the CR blocks in FRCRN, we use $C^\prime=128$ to keep the same number of feature maps.  The complex BN and the LeakyReLU follow \cite{Trabelsi2018}. 

The output of the Conv2d layer after the complex BN and the LeakyReLU function is sent to the CFSMN layer. The reason that we apply FSMN instead of using LSTM is because that FSMN not only achieves competitive performance compared to LSTM, but also requires only about a quarter of the parameters of LSTM \cite{Zhang2018}. The operations of CFSMN used in FRCRN is described as follows. 
Consider the real part $U_r$ of the feature map $U$ and permute it from $C^\prime \times T \times F^{\prime\prime}$ to $T \times F^{\prime\prime} \times C^\prime$. For the current frame $t$ of $U_r \in \mathbb{R}^{T \times F^{\prime\prime} \times C^\prime}$, we form the frequency sequence $\mathbf{S}_{r}(t)\in \mathbb{R}^{F^{\prime\prime}\times C^\prime}=\{\mathbf{s}_{f_1}(t), \mathbf{s}_{f_2}(t),\cdots,\mathbf{s}_{F^{\prime\prime}}(t)\in \mathbb{R}^{C^\prime \times 1}\}$. Applying real cell of CFSMN on the sequence $\mathbf{S}_r(t)$, the output of the $l$th component takes the following form:
\begin{align}
&\mathbf{h}_{f_i}^l=\delta(\mathbf{W}_{f_i}^l\mathbf{s}_{f_i}^{l-1}+\mathbf{b}_{f_i}^l) \\
&\mathbf{p}_{f_i}^l=\mathbf{V}_{f_i}^l\mathbf{h}_{f_i}^{l}+\mathbf{v}_{f_i}^l \\
&\mathbf{s}_{f_i}^{l}=\mathbf{s}_{f_i}^{l-1}+\mathbf{p}_{f_i}^l+\sum_{\tau=0}^{N_L}{\mathbf{a}_{\tau}^l\odot\mathbf{p}_{f_i-\tau}^l}+\sum_{\kappa=0}^{N_R}{\mathbf{c}_{\kappa}^l\odot\mathbf{p}_{f_i+\kappa}^l}
\end{align}
Here, $f_i=f_1,f_2,\cdots,F^{\prime\prime}$, and $t$ is omitted for simplicity in expression. $\delta$ stands for ReLU function. $N_L$ and $N_R$ denotes the look-back and lookahead orders of the $l$th memory block, respectively. We set $N_L=20$ and $N_R=0$ to only look backwards in time for all sequences in our experiments. Since we use a single CFSMN layer in the CR block, the value of $l$ is $1$. $\mathbf{s}_{f_i}^0$ is equivalent to $\mathbf{s}_{f_i}$ and $\mathbf{s}_{f_i}^1$ is the output. For the imaginary part, we apply an imaginary cell of CFSMN with the same operation as the real cell. The CFSMN output $\mathbf{S}_\mathrm{out}\in \mathbb{C}^{F^{\prime\prime}\times C^\prime}$ can be defined as:
\begin{align}
\mathbf{S}_\mathrm{out}&=\mathrm{FSMN}_r(\mathbf{S}_r)-\mathrm{FSMN}_i(\mathbf{S}_i) \nonumber \\
&+j(\mathrm{FSMN}_r(\mathbf{S}_i)+\mathrm{FSMN}_i(\mathbf{S}_r)) 
\end{align}
where $\mathrm{FSMN}_r$ and $\mathrm{FSMN}_i$ represent real and imaginary cells of CFSMN. $\mathbf{S}_r$ and $\mathbf{S}_i$ denote real and imaginary parts of the frequency sequence. 

In the recurrent module, the real part of CFSMN input is reshaped to $U_r \in \mathbb{R}^{T \times H}$ with $H=F^{\prime\prime} \times C^\prime$. We then form a time sequence $\mathbf{Q}_r=\{\mathbf{q}_{t_1}, \mathbf{q}_{t_2}, \cdots, \mathbf{q}_{T} \in\mathbb{R}^{H\times 1}\}$. The operations of Eqn. (2) to (4) are applied on $\mathbf{Q}_r$ to model time dynamics of $U_r$. The outputs of CFSMN follows Eqn. (5). 

\subsection{Joint loss function}
The time domain loss function SI-SNR \cite{Lv2021} has been commonly used as an evaluation metric in noise suppression. It is a signal-level loss and directly performs on the signal itself. To further guide the learning of cIRM estimation, we also considered the mean squared error (MSE) losses of the real ${M}_r$ and imaginary ${M}_i$ estimates of cIRM \cite{Zhao2021}. Specifically, we optimize the FRCRN model by the following joint loss function:
\begin{equation}
\mathcal{L}(y,\hat{y}) = \mathcal{L}_{SI-SNR}(y,\hat{y})+\lambda\mathcal{L}_{Mask}(M,\hat{M}),
\end{equation}
where $\mathcal{L}_{SI-SNR}(y,\hat{y})$ is the SI-SNR loss defined as \cite{Lv2021}:
\begin{equation}
\mathcal{L}(y,\hat{y}) = \mathrm{10log_{10}}\bigg(\frac{\|y_\mathrm{target}\|^2_2}{||e_\mathrm{noise}||^2_2}\bigg)
\end{equation}
with $y_\mathrm{target}=(\langle\hat{y},y\rangle\cdot y)/||y||^2_2$ and $e_\mathrm{noise}=\hat{y}-y_\mathrm{target}$. Here, $\langle\cdot,\cdot\rangle$ denotes dot product and $||\cdot||_2$ is the L2 norm.
The mask loss $\mathcal{L}(M,\hat{M})$ is defined as: 
\begin{equation}
\mathcal{L}(M,\hat{M}) = \sum_{t,f}[(\hat{M}_r-M_r)^2+(\hat{M}_i-M_i)^2]
\end{equation}
In this work, we give equal weights to the signal estimation and the cIRM estimation by setting the factor $\lambda$ to 1.

\begin{table}
\center
\footnotesize
\caption{Objective evaluation results of various models on WSJ0 dataset. }
\begin{tabular}{lcccc}
\specialrule{.1em}{.05em}{.05em}
\multirow{2}{*}{Model} &\multirow{2}{*}{Para.(M)} & \multicolumn{3}{c}{Evaluation Metrics}         \\
\cline{3-5}
                                 &       & PESQ   & STOI   & SI-SNR     \\ \hline
Noisy                            & -     & 1.97   & 87.83  & 6.28          \\ \hline
DCCRN \cite{Hu2020}          	  & 3.7   & 3.17   & 95.80  & 17.71          \\ 
PHASEN \cite{Yin2020}            & 8.8   & 3.38   & 96.58  & 18.33         \\ \hline
FRCRN-Lite                       & 2.1   & 3.21   & 96.07  & 18.11          \\ 
FRCRN                            & 6.9   & \textbf{3.62}   & \textbf{98.24}  & \textbf{21.33}  \\
w/o CFSMN in CRED                  & 6.0   & 3.42   & 97.36  & 20.10        \\ 
w/o Attention Block                & 5.8   & 3.31   & 96.70  & 18.40         \\
w/o Recurrent Module               & 5.7   & 3.23   & 96.29  & 18.35        \\ \hline
\specialrule{.1em}{.05em}{.05em}
\end{tabular}
\end{table}

\section{Experiments}
\label{sec:typestyle}
\subsection{Evaluation on wideband speech signal}
As most of the previous models are tested on wideband speech enhancement, we first evaluate our proposed FRCRN model on  speech signals of 16kHz. For wideband speech inputs, we use the window length of 20ms  and the frame shift of 10ms, respectively. The total latency of the wideband FRCRN model is limited to 30ms. The STFT length is set to 640 with zero-paddings for an increased spectral resolution. The number of input channel of the model is 1.  The number of CR blocks in the encoder and the decoder of CRED is 6. The time-recurrent module uses 2 stacked CFSMN layers. Each CR block uses 128 channels for the convolution operation and 128 units for the real part and the imaginary part of CFSMN, respectively.  
We use PyTorch and Adam optimizer with a batch size of 12 for model optimization. The training starts with an initial learning rate of 1e-3 with a decay of 0.98. Our trainings are ended at around 120 epochs by early stopping. 
\subsubsection{Ablation study on WSJ0 dataset}
We first conduct an ablation study to evaluate the full FRCRN model as well as the contribution of each sub-module. To show the model capability under a low complexity, we also include a lite FRCRN (FRCRN-Lite) model by setting the convolution channels $C^\prime$ and the CFSMN cell size to 64, which has three times less trainable parameters compared to the full FRCRN model. We choose a CRN-based DCCRN \cite{Hu2020} model and a two-stream network, PHASEN \cite{Yin2020}, as baselines. Both DCCRN and PHASEN belong to the complex-domain family and estimate the magnitude and phase simultaneously. Our implementation for the baseline models follows the best configuration mentioned in the literature. In this evaluation, 50 hours of clean speech from 131 speakers were selected from the WSJ0 corpus \cite{Garofolo1993} and 50 hours of noise were selected from  RNNoise. 40 hours of clean speech and 40 hours of noise were used for training and validation, and the rest for the test set. A wide range of SNRs between 0 dB and 10 dB were included in the test set. The reverberation corruption is not considered for ease of comparison. Three objective metrics including PESQ, STOI, and SI-SNR were used for evaluation. 

The results are shown in Table 1.  From the evaluation results, we can see that the proposed full FRCRN model outperforms the baselines with a large margin on all the evaluation metrics. The FRCRN-Lite has a lower complexity than DCCRN, but still provide a competitive performance. Our ablation study is conducted by progressively (1) removing CFSMN layers from CRED; (2) removing the attention block CCBAM; (3) removing the Recurrent Module. The trend of performance degradation as shows that all components are contributive, especially the frequency recurrent CFMN layers in CRED.  This verifies benefits of the proposed CRED by leveraging on both the spatio-temporal patterns and the long-range frequency dependencies.
\begin{table}
\center
\footnotesize
\caption{Comparison with other SOTA models on the DNS-2020 non-blind test set. }
\begin{tabular}{lccccc}
\specialrule{.1em}{.05em}{.05em}
\multirow{2}{*}{Model} &\multirow{2}{*}{Year} & \multicolumn{4}{c}{Evaluation Metrics}         \\
\cline{3-6}
                   &    & WB-PESQ   & PESQ   & STOI   & SI-SNR  \\ \hline
Noisy              &    & 1.58      & 2.45   & 91.52   & 9.07     \\ \hline
NSNet \cite{Reddy2020}         	  &2020   & 2.15      & 2.87   & 94.47   & 15.61      \\ 
DTLN \cite{Westhausen2020}       &2020    &  -        & 3.04   & 94.76   & 16.34     \\
DCCRN \cite{Hu2020}   &2020    &  -        & 3.27   &   -     &   -     \\
FullSubNet \cite{Hao2020}         &2021    & 2.78      & 3.31   & 96.11   & 17.29     \\ 
TRU-Net \cite{Choi2021}            &2021    & 2.86      & 3.36   & 96.32   & 17.55     \\
DCCRN+ \cite{Lv2021}            &2021   &  -        & 3.33   &   -     &   -     \\
CTS-Net \cite{Li22021}           &2021   & 2.94      & 3.42   & 96.66   & 17.99     \\ 
GaGNet \cite{Li2021}           &2021    & 3.17      & 3.56   & 97.13   & 18.91    \\ \hline
FRCRN  &2021    & \textbf{3.23}      & \textbf{3.60}   & \textbf{97.69}   & \textbf{19.78}     \\ \hline
\specialrule{.1em}{.05em}{.05em}
\end{tabular}
\end{table}

\begin{table}
\center
\footnotesize
\caption{Comparison with other SOTA models on the VoiceBank+Demand test set. }
\begin{tabular}{lccccc}
\specialrule{.1em}{.05em}{.05em}
\multirow{2}{*}{Model} &\multirow{2}{*}{Year}& \multicolumn{4}{c}{Evaluation Metrics}         \\
\cline{3-6}
                &       & WB-PESQ   & CSIG   & CBAK   & COVL  \\ \hline
Noisy           &       & 1.97      & 3.35   & 2.44   & 2.63     \\ \hline
SEGAN \cite{Pascual2017}          &2017	   & 2.16      & 3.48   & 2.94   & 2.80      \\ 
HiFi-GAN \cite{Su2020}        &2020       & 2.94      & 4.07   & 3.07   & 3.49     \\
GaGNet \cite{Li2021}         &2021       & 2.94      & 4.26   & 3.45   & 3.59     \\ 
PHASEN \cite{Yin2020}         &2020       & 2.99      & 4.21   & 3.55   & 3.62     \\
DEMUCS \cite{Pascual22017}         &2021      & 3.07      & 4.31   & 3.40   & 3.63     \\ 
MetricGAN+ \cite{Fu2021}     &2021       & 3.15      & 4.14   & 3.16   & 3.64    \\ 
PERL-AE \cite{Kataria2021}        &2021       & 3.17      & \textbf{4.43}   & 3.53   & \textbf{3.83}    \\ \hline
FRCRN &2021      & \textbf{3.21}      & 4.23   & \textbf{3.64}   & 3.73     \\ \hline
\specialrule{.1em}{.05em}{.05em}
\end{tabular}
\end{table}
\subsubsection{Evaluation on two benchmarks}
We further conducted evaluations on two popular benchmarks: DNS-2020 dataset \cite{Reddy2020} and VoiceBank+Demand dataset \cite{Botinhao2016}. Both datasets are wideband speech signals. The DNS-2020 dataset contains 500 hours of clean speech, 65K noise clips and 80K RIR clips. We generate a total of 3K hours noisy-clean pairs for training and development, of which 30\% of clean speech is convolved with the RIR clips. During training data generation, the SNR is randomly selected between 0 and 15 dB. The non-blind synthetic test set is adopted for objective evaluation with wideband PESQ (WB-PESQ) included. The VoiceBank+Demand dataset is a relatively small evaluation dataset, which consists of 11,572 clean-noisy training pairs from 28 speakers and 824 pairs from another 2 speakers for testing. Four metrics are utilized, namely WB-PESQ, CSIG, CBAK, and COVL \cite{Loizou2007}. 

Tables 2 and 3 show the performance comparisons with the published results of the SOTA models.  One can find that our proposed FRCRN model outperforms the previous SOTA methods on most of the evaluation metrics, except that PERL-AE leveraging a large-scale pre-trained model performs better on CSIG and COVL. Our processed samples are available at the repository \footnote{https://github.com/alibabasglab/FRCRN}.
\begin{table}
\center
\footnotesize
\caption{Evaluation on DNS-2022 development set. }
\begin{tabular}{lcccc}
\specialrule{.1em}{.05em}{.05em}
\multirow{2}{*}{Model} &\multirow{2}{*}{WAcc}& \multicolumn{3}{c}{DNSMOS}         \\
\cline{3-5}
                          &         & SIG       & BAK   & OVRL  \\ \hline
Noisy                     & 0.631   & 3.866     & 2.976   & 3.075   \\ 
NSNet2 \cite{Dubey2022} & 0.544   & 3.606     & 4.074   & 3.281   \\ \hline
FRCRN                     & \textbf{0.641}   & \textbf{3.999}     & \textbf{4.097}   & \textbf{3.545}   \\ \hline
\specialrule{.1em}{.05em}{.05em}
\end{tabular}
\end{table}
\begin{table}
\center
\footnotesize
\caption{Evaluation on DNS-2022 blind test set. }
\begin{tabular}{lccccc}
\specialrule{.1em}{.05em}{.05em}
\multirow{2}{*}{Model} &\multirow{2}{*}{WAcc} & \multicolumn{3}{c}{MOS} &\multirow{2}{*}{Final Score}    \\
\cline{3-5}
                          &         & SIG       & BAK   & OVRL  &\\ \hline
Noisy                     & \textbf{0.72}   & \textbf{4.29}     & 2.15   & 2.63   & 0.56\\ 
NSNet2 \cite{Dubey2022} & 0.63   & 3.62     & 3.93   & 3.26  & 0.60\\ \hline
FRCRN (Team14)          & 0.69   & 4.26     & \textbf{4.27}   & \textbf{3.89}   &\textbf{0.70}\\ \hline
\specialrule{.1em}{.05em}{.05em}
\end{tabular}
\end{table}
\subsection{Evaluation on fullband speech signal}
We further extended the evaluation of our proposed FRCRN model on the fullband DNS-2022 dataset \cite{Dubey2022}. For the training setup, we keep the same window length of 20ms and frame shift of 10ms. Therefore, the algorithm delay is 30ms. To deal with the speech signal sampled at 48kHz, we make the following changes. We increase the STFT length to 1920 and increase the input spectrogram channel from 1 to 3. We assign the frequency bins from 1 to 641 to the 1st channel, frequency bins from 641 to 1282 to the 2nd channel, and frequency bins from 1282 to 1921 to the 3rd channel. The dimension of the target cIRM increases from 641 in wideband speech to 1921 in fullband speech. All the other training setup is the same as described in Section 3.1. The total trainable parameters of the fullband FRCRN model is 10.27 million, and the number of multiply-accumulate operations (MACS) is 12.30 GMACS per second. 

Using the fullband data provided by the DNS-2022, we generated a total of 3K hours noisy-clean pairs and 30\% of which are reverberant speech.
Table 4 shows the evaluation results on the DNS-2022 development set. We use both DNSMOS P.835 \cite{reddy2020dnsmos} and Word Accuracy (WAcc) metrics provided by the organizer for our evaluation. The DNSMOS score measures speech quality (SIG), background noise quality (BAK), and overall audio quality (OVRL), respectively. The WAcc score measures model impact on speech recognition performance. As can be seen from Table 4 our model achieves better results on all metrics compared to the baseline. We use the same model to enhance the blind test set and Table 5 shows the DNS-2022 P.835 \cite{naderi2021crowdsourcing} subjective evaluation results and WAcc results. Our model has a consistent better performance than the baseline with lighter degradation on speech quality and WAcc. Our submission ranked top 2 in final score in the non-personalized track. 
\section{Conclusions}
In this work, we proposed a convolutional recurrent encoder-decoder structure (CRED) to boost feature representation based on frequency recurrence. The frequency recurrence is applied on the 3D convolutional feature maps along the frequency axis and is efficiently realized by a feedforward sequential memory network. The FRCRN model utilizes CRED to capture long-range frequency correlations and the time-recurrent module to capture the temporal dynamics. We implemented FRCRN in the complex-valued domain and used a joint loss function for optimization. Our FRCRN model achieved SOTA performance on wideband benchmarks and 2nd place in the fullband non-personalized track in the ICASSP 2022 DNS challenge. 
\bibliographystyle{IEEEbib}
\bibliography{refs}

\begin{thebibliography}{10}

\bibitem{Gao2016}
T.~Gao, J.~Du, L.~R. Dai, and C.~H. Lee,
\newblock ``{SNR}-based progressive learning of deep neural network for speech
  enhancement,''
\newblock in {\em Interspeech}, 2016.

\bibitem{Gao2018}
T.~Gao, J.~Du, L.~R. Dai, and C.~H. Lee,
\newblock ``Densely connected progressive learning for lstm-based speech
  enhancement,''
\newblock in {\em IEEE ICASSP}, 2018, pp. 5054--5058.

\bibitem{Park2017}
S.~R. Park and J.~W. Lee,
\newblock ``A fully convolutional neural network for speech enhancement,''
\newblock in {\em Interspeech}, 2017.

\bibitem{Tan2018}
K.~Tan and D.~Wang,
\newblock ``A convolutional recurrent neural network for real-time speech
  enhancement,''
\newblock in {\em Proc. Interspeech}, 2018.

\bibitem{Zhao2018}
H.~Zhao, S.~Zarar, I.~Tashev, and C.-H. Lee,
\newblock ``Convolutionalrecurrent neural networks for speech enhancement,''
\newblock in {\em IEEE ICASSP}, 2018, p. 2401–2405.

\bibitem{Hu2020}
Y.~Hu, Y.~Liu, S.~Lv, M.~Xing, S.~Zhang, Y.~Fu, J.~Wu, B.~Zhang, and L.~Xie,
\newblock ``{DCCRN}: Deep complex convolution recurrent network for phase-aware
  speech enhancement,''
\newblock in {\em Proc. Interspeech}, 2020.

\bibitem{Zhao2021}
S.~Zhao, T.~H. Nguyen, and B.~Ma,
\newblock ``Monaural speech enhancement with complex convolutional block
  attention module and joint time frequency losses,''
\newblock in {\em IEEE ICASSP}, 2021.

\bibitem{Lv2021}
S.~Lv, Y.~Hu, S.~Zhang, and L.~Xie,
\newblock ``{DCCRN}+: Channel-wise subband dccrn with snr estimation for speech
  enhancement,''
\newblock {\em arXiv preprint arXiv:2106.08672}, 2021.

\bibitem{Zhang2018}
S.~Zhang, M.~Lei, Z.~Yan, and L.~Dai,
\newblock ``Deep-fsmn for large vocabulary continuous speech recognition,''
\newblock {\em arXiv preprint arXiv:1803.05030}, 2018.

\bibitem{Williamson2015}
D.~S. Williamson, Y.~Wang, and D.~Wang,
\newblock ``Complex ratio masking for monaural speech separation,''
\newblock {\em IEEE/ACM transactions on audio, speech, and language
  processing}, vol. 24, no. 3, pp. 483– 492, 2015.

\bibitem{Reddy2020}
C.~K. Reddy, V.~Gopal, R.~Cutler, E.~Beyrami, R.~Cheng, H.~Dubey,
  S.~Matusevych, R.~Aichner, A.~Aazami, and S.~Braun {\it et al}.,
\newblock ``The interspeech 2020 deep noise suppression challenge: Datasets,
  subjective testing framework, and challenge results,''
\newblock {\em arXiv preprint arXiv:2005.13981}, 2020.

\bibitem{Botinhao2016}
C.Valentini-Botinhao, X.Wang, S.Takaki, and J.Yamagishi,
\newblock ``Investigating {RN}n-based speech enhancement methods for
  noise-robust {T}ext-to-{S}peech,''
\newblock in {\em Proc. SSW}, 2016, p. 146–152.

\bibitem{Dubey2022}
H.~Dubey, V.~Gopal, R.~Cutler, A.~Aazami, S.~Matusevych, S.~Braun, S.~E.
  Eskimez, Ma. Thakker, T.~Yoshioka, H.~Gamper, and R.~Aichner,
\newblock ``Icassp 2022 deep noise suppression challenge,''
\newblock in {\em IEEE ICASSP 2022}, 2022.

\bibitem{Trabelsi2018}
C.~Trabelsi, O.~Bilaniuk, Y.~Zhang, D.~Serdyuk, S.~Subramanian, J.~F. Santos,
  S.~Mehri, N.~Rostamzadeh, Y.~Bengio, and C.~J. Pal.,
\newblock ``Deep complex networks,''
\newblock in {\em International Conference on Learning Representations}, 2018.

\bibitem{Yin2020}
D.~Yin, C.~Luo, Z.~Xiong, and W.~Zeng,
\newblock ``Phasen: A phase-and- harmonics-aware speech enhancement network,''
\newblock in {\em Proc. AAAI}, 2020, vol.~34, p. 9458–9465.

\bibitem{Garofolo1993}
J.~Garofolo, D.~Graff, D.~Paul, and D.~Pallett,
\newblock ``Csr-i (wsj0) complete ldc93s6a,''
\newblock {\em Web Download. Philadelphia: Linguistic Data Consortium}, vol.
  83, 1993.

\bibitem{Westhausen2020}
N.~L. Westhausen and B.~T. Meyer,
\newblock ``Dual-signal transformation lstm network for real-time noise
  suppression,''
\newblock in {\em Proc. Interspeech}, 2020, p. 2477–2481.

\bibitem{Hao2020}
X.~Hao, X.~Su, R.~Horaud, and X.~Li,
\newblock ``{F}ull{S}ubnet: A full-band and sub-band fusion model for real-time
  single-channel speech enhancement,''
\newblock {\em arXiv preprint arXiv:2010.15508}, 2020.

\bibitem{Choi2021}
H.-S. Choi, S.~Park, J.~H. Lee, H.~Heo, D.~Jeon, and K.~Lee,
\newblock ``Real-time denoising and dereverberation with tiny recurrent
  u-net,''
\newblock {\em arXiv preprint arXiv:2102.03207}, 2021.

\bibitem{Li22021}
A.~Li, W.~Liu, C.~Zheng, and X.~Li,
\newblock ``Two heads are better than one: A two-stage complex spectral mapping
  approach for monaural speech enhancement,''
\newblock {\em IEEE/ACM Trans. Audio. Speech, Lang. Process.}, vol. 29, pp.
  1829–1843, 2021.

\bibitem{Li2021}
A.~Li, C.~Zheng, and X.~Li L.~Zhang,
\newblock ``{G}lance and {G}aze: {A} collaborative learning framework for
  single-channel speech enhancement,''
\newblock {\em arXiv preprint arXiv:2106.11789}, 2021.

\bibitem{Pascual2017}
S.~Pascual, A.~Bonafonte, and J.~Serra,
\newblock ``Segan: Speech enhancement generative adversarial network,''
\newblock in {\em Proc. Interspeech}, 2017, p. 3642–3646.

\bibitem{Su2020}
J.~Su, Z.~Jin, and A.~Finkelstein,
\newblock ``{H}i{F}i-{GAN}: High-fidelity denoising and dereverberation based
  on speech deep features in adversarial networks,''
\newblock {\em arXiv preprint arXiv:2006.05694}, 2020.

\bibitem{Pascual22017}
A.~Defossez, G.~Synnaeve, and Y.~Adi,
\newblock ``Real time speech enhancement in the waveform domain,''
\newblock in {\em Proc. Interspeech}, 2020, p. 3291–3295.

\bibitem{Fu2021}
S.-W. Fu, C.~Yu, T.-A. Hsieh, P.~Plantinga, M.~Ravanelli, X.~Lu, and Y.~Tsao,
\newblock ``{M}etric{GAN}+: An improved version of {M}etric{GAN} for speech
  enhancement,''
\newblock {\em arXiv preprint arXiv:2104.03538}, 2021.

\bibitem{Kataria2021}
S.~Kataria, J.~Villalba, and N.~Dehak,
\newblock ``Perceptual loss based speech denoising with an ensemble of audio
  pattern recognition and self-supervised models,''
\newblock in {\em IEEE ICASSP}, 2021.

\bibitem{Loizou2007}
Y.~Hu and P.C. Loizou,
\newblock ``Evaluation of objective quality measures for speech enhancement,''
\newblock {\em IEEE/ACM Trans. Audio. Speech, Lang. Process.}, vol. 16, no. 1,
  pp. 229–238, 2007.

\bibitem{reddy2020dnsmos}
Chandan~KA Reddy, Vishak Gopal, and Ross Cutler,
\newblock ``Dnsmos: A non-intrusive perceptual objective speech quality metric
  to evaluate noise suppressors,''
\newblock in {\em ICASSP}, 2020.

\bibitem{naderi2021crowdsourcing}
Babak Naderi and Ross Cutler,
\newblock ``Subjective evaluation of noise suppression algorithms in
  crowdsourcing,''
\newblock in {\em INTERSPEECH}, 2021.

\end{thebibliography}

\end{document}